\documentclass[conference]{IEEEtran}
\IEEEoverridecommandlockouts
\usepackage{cite}
\usepackage{amsmath,amssymb,amsfonts}
\usepackage{algorithmic}
\usepackage{graphicx}
\usepackage{textcomp}
\usepackage{xcolor}
\def\BibTeX{{\rm B\kern-.05em{\sc i\kern-.025em b}\kern-.08em
    T\kern-.1667em\lower.7ex\hbox{E}\kern-.125emX}}
\usepackage{bbding}
\usepackage{algorithm}
\usepackage{epstopdf}
\begin{document}

\title{{Reduced-Complexity Cross-Domain Iterative Detection for OTFS Modulation via Delay-Doppler Decoupling} \vspace{-0.25cm}}
\author{
	\IEEEauthorblockN{Mengmeng Liu\IEEEauthorrefmark{1},
		Shuangyang Li\IEEEauthorrefmark{2},		
		Baoming Bai\IEEEauthorrefmark{1}, 
		and
		Giuseppe Caire\IEEEauthorrefmark{2}
	}
	\IEEEauthorblockA{\IEEEauthorrefmark{1}State Key Lab. of ISN,
		Xidian University, Xi'an, China}
	\IEEEauthorblockA{\IEEEauthorrefmark{2}Technical University of Berlin, Germany}
	 \IEEEauthorblockA{Email: mmengliu@stu.xidian.edu.cn, shuangyang.li@tu-berlin.de, bmbai@mail.xidian.edu.cn, caire@tu-berlin.de }
	\vspace{-1cm}
}
\maketitle
\begin{abstract}
	In this paper, a reduced-complexity cross-domain iterative detection for orthogonal time frequency space (OTFS) modulation is proposed, which exploits channel properties in both time and delay-Doppler domains. 
	Specifically, we first show that in the time domain effective channel, the path delay only introduces interference among samples in adjacent time slots, while the Doppler becomes a phase term that does not affect the channel sparsity.
	This ``band-limited'' matrix structure motivates us to apply a reduced-size linear minimum mean square error (LMMSE) filter to eliminate the effect of delay in the time domain, while exploiting the cross-domain iteration for minimizing the effect of Doppler by noticing that the time and Doppler are a pair of Fourier dual. 
	The state (MSE) evolution was derived and compared with bounds to verify the effectiveness of the proposed scheme.
	Simulation results demonstrate that the proposed scheme achieves almost the same error performance as the optimal detection, but only requires a reduced complexity.
\end{abstract}


\section{Introduction}

Future wireless networks are envisioned to accommodate many emerging applications, such as low-earth orbit (LEO) satellites and unmanned aerial vehicles (UAVs), where the signal is inevitably transmitted over complex and challenging channel scenarios.
The recently proposed orthogonal time frequency space (OTFS) modulation has shown to be a good solution to signal transmissions over such channels \cite{Hadani-2017WCNC,L.Gaudio-2022TWC}.
The information symbols in OTFS systems are multiplexed in the delay-Doppler (DD) domain, leading to the full exploration of appealing DD domain channel properties, including quasi-static, separable, and sparse \cite{Z.Wei-2021WCMagazine}, which in return facilitates the design of channel estimation and equalization.
More importantly, OTFS can potentially achieve the full channel diversity \cite{Hadani-2017WCNC}, which ensures better performance robustness compared to currently deployed orthogonal frequency division multiplexing (OFDM) over challenging transmission scenarios~\cite{Z.Wei-2021WCMagazine}.

The promising performance of OTFS relies on advanced equalization. However, optimal maximum-likelihood sequence estimation (MLSE) usually requires prohibitively high detection complexity and cannot be directly applied to practical systems. Thus, the design of reduced-complexity detection for OTFS has acquired much attention.
For instance, a message passing (MP) algorithm based on maximum \emph{a posteriori} probability (MAP) was proposed in \cite{P.Raviteja-2018TWC}, where inter-symbol interference (ISI) is Gaussian-approximated to reduce detection complexity.
An improved MP detector for OTFS was proposed in \cite{L.Gaudio-2020TWC}, which can avoid $4$-cycles in the factor graph.
Furthermore, many improved detection algorithms based on MP were proposed, such as hybrid MAP and parallel interference cancellation \cite{S.Li-2021TVT-MAP-PIC} and Gaussian approximate MP \cite{Xiang-2021TVT-GAMP}.
Note that most OTFS detection schemes, including the aforementioned ones, operate in the DD domain.
However, when the time and frequency resources for OTFS are limited, the DD domain channel matrix could be dense due to insufficient resolution of delay and Doppler, and consequently, DD domain detection may suffer from high detection complexity.
The cross-domain iterative detection proposed in \cite{S.Li-2022TWC-Crossdomain} was a preliminary attempt to solve this issue by considering the detection in both time and DD domains via iterative processing, which can achieve almost the same error performance as ML detection even in the presence of fractional Doppler shifts.
The cross-domain iterative detection is motivated by the unitary transformation between the time and DD domains, ensuring that the detection error in one domain is principally orthogonal to that in the other domain. Thus, it allows cross-domain iterations for signal detection without introducing error propagation.
However, \cite{S.Li-2022TWC-Crossdomain} applied a full-size linear minimum mean square error (LMMSE) filter in the time domain, which, as we will show later, does not fully exploit the advantages of cross-domain iteration.

In this paper, we propose a novel cross-domain iterative detection for OTFS with reduced complexity. The major motivation is that the effects of delay and Doppler can be decoupled and thus can be treated separately. We first review the properties of the time domain OTFS effective channel matrix, where we show that the path delay only introduces interference among samples in adjacent time slots, while the Doppler behaves as a phase term that does not affect the channel sparsity. Based on the ``band-limited'' matrix structure, we propose a reduced-size estimator in the time domain to eliminate the effect of delay, while relying on cross-domain iterations to minimize the effect of Doppler. Such a scheme is motivated by the fact that the time and Doppler are a pair of Fourier dual and therefore the effect of Doppler can be minimized by iteratively exchanging the extrinsic information between the time and DD domains.
As a result, our proposed detection enjoys much lower complexity while maintaining the promising error performance of cross-domain iterative detection. The effectiveness of the proposed scheme was verified by the derived state (MSE) evolution, which is also compared with theoretical performance bounds. The numerical results agree with our analysis and demonstrate a near-optimal error performance.

\textit{Notations}: 
${{\bf{F}}_N}$ and ${\bf{F}}_N^{\rm{H}}$ denote the discrete Fourier transform (DFT) matrix and inverse DFT (IDFT) matrix of size $N \times N$, respectively;  
${\bf I}_{N}$ denotes the identity matrix of size $N \times N$;
$\otimes$ denotes the Kronecker product operator; 
$\left(  \cdot  \right)^{\rm H}$ denotes the Hermitian transpose;
$\left(  \cdot  \right)^{\rm T}$ denotes the transpose;
${\rm{diag}} \left\{ \cdot \right\}$ denotes the diagonal matrix;
$\delta \left( \cdot \right)$ is the Dirac delta function.

\section{System Model}
\subsection{Backgrounds on OTFS Transmissions}
\begin{figure}[htp]
	\centering
	\includegraphics[width=3.5in,height=0.75in]{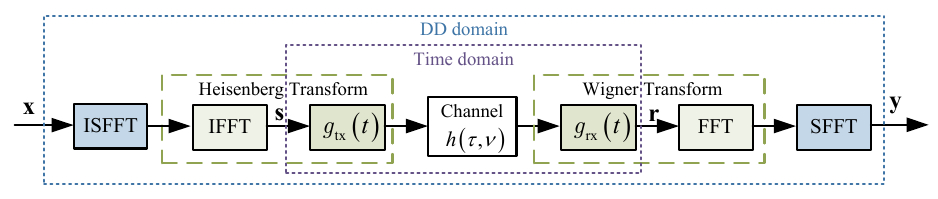}
	\vspace{-0.8cm}
	\caption{The transceiver structure of OTFS transmissions.}\label{Fig-Transmitter}
\end{figure}

\vspace{-0.05cm}
Without loss of generality, let us consider the transceiver structure of OTFS transmissions in Fig. \ref{Fig-Transmitter}.
Let $M$ be the number of delay bins (sub-carriers) and $N$ be the number of Doppler bins (time slots). Let $T$ be the duration of each time slot, and correspondingly, the sub-carrier spacing is $1/T$. 
A length-$MN$ DD domain information symbol vector $\bf x$ is passed through an OTFS modulator, which performs the inverse symplectic finite Fourier transform (ISFFT) and the Heisenberg transform. 
The resultant discrete time domain transmitted signal $\bf s$ is denoted as
\begin{equation}\label{s=x}
\setlength{\abovedisplayskip}{4pt} 
\setlength{\belowdisplayskip}{4pt}
{\bf{s}} = \left( {{\bf{F}}_N^{\rm{H}} \otimes {{\bf{I}}_M}} \right){\bf{x}}.
\end{equation}


Consider a path-$P$ linear time-varying (LTV) wireless channel given by
\begin{equation}\label{h-tao-viu}
\setlength{\abovedisplayskip}{3pt} 
\setlength{\belowdisplayskip}{4pt}
h\left( {\tau ,\nu } \right) = \sum\limits_{p = 1}^P {{h_p}\delta \left( {\tau  - {\tau _p}} \right)\delta \left( {\nu  - {\nu _p}} \right)}, 
\end{equation}
where $h_p$ is the fading coefficient for the $p$-th path, following a complex Gaussian distribution with zero mean and variance ${1 \mathord{\left/{\vphantom {1 {\left( {2P} \right)}}} \right. \kern-\nulldelimiterspace} {\left( {2P} \right)}}$ per real dimension (uniform power profile); and $\tau _p \in \left[0,T \right)$ and $\nu _p \in \left[0,1/T \right)$ represent the delay and Doppler shifts associated with the $p$-th path, respectively. Particularly, we consider the discretized delay and Doppler indices defined by ${l_p} = {{\tau_p}M}/T$ and ${k_p}+{\kappa _p} = {{\nu _p}NT}$,
where $0 \le l_p \le M-1$ and $0 \le k_p \le N-1$ are the corresponding integer indices of delay and Doppler for the $p$-th path, while ${\kappa _p} \in \left(-0.5,0.5\right]$ represents the fractional contribution of the Doppler shift. In this paper, we focus only on the integer delay case, which only asymptotically holds with a sufficiently large signal bandwidth. 
In fact, our proposed scheme can also work in the fractional delay case and more details can be found in our journal paper.

Assume that ${g_{\rm{tx}}}\left(t\right)$ and ${g_{\rm{rx}}}\left(t\right)$ are rectangular pulses, and a reduced-cyclic-prefix (reduced-CP) structure is applied to the system, the time domain vectorized input-output relation for OTFS transmissions after CP removal can be given by
\begin{equation}\label{r-Hs}
\setlength{\abovedisplayskip}{4pt} 
\setlength{\belowdisplayskip}{4pt}
{\bf{r}} = {{\bf{H}}_{\rm T}} {\bf{s}} + {\bf{n}},
\end{equation}
where $\bf{r}$ is the time domain received vector of length $MN$, and $\bf{n}$ is the additive white Gaussian noise vector with zero mean and one-sided power spectral density of $N_0$. According to \cite{S.Li-2022-THP-OTFS}, the time domain effective channel matrix ${{\bf{H}}_{\rm T}}$ can be expressed as
\begin{figure}[htp]
	\centering
	\vspace{-0.5cm}
	\includegraphics[width=3.5in,height=1.3in]{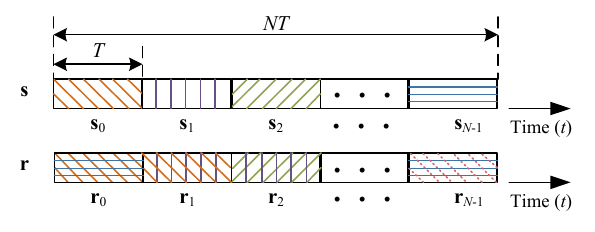}
	\vspace{-0.8cm}
	\caption{Brief illustration of the interference pattern between the time domain transmitted vector $\bf s$ and the received vector $\bf r$.}\label{Fig-Time-domain-input-output}
	\vspace{-0.5cm}
\end{figure}
\begin{equation}\label{H-T}
\setlength{\abovedisplayskip}{4pt} 
\setlength{\belowdisplayskip}{4pt}
{\bf{H}}_{\rm T} = \sum\limits_{p = 1}^P {{h_p}{e^{ - j2\pi \frac{{{k_p} + {\kappa _p}}}{{NM}}{l_p}}}{{\bf{\Delta }}^{{k_p} + {\kappa _p}}}{{\bf{\Pi }}^{{l_p}}}} ,
\end{equation}
where, ${\bf{\Delta }}= {\rm{diag}}\left\{ {{\alpha ^0},{\alpha ^1},\ldots,{\alpha ^{MN - 1}}} \right\}$ is a diagonal matrix with $\alpha  \buildrel \Delta \over = {e^{\frac{{j2\pi }}{{MN}}}}$ characterizing the Doppler influence, and  ${\bf{\Pi }} = {\rm circ}\left\{ {\left[ {0,1,0, \ldots ,0} \right]_{MN \times 1}^{\rm{T}}} \right\}$ is the permutation matrix (forward cyclic shift) characterizing the delay influence.
Finally, the DD domain received symbol vector $\bf y$ can be obtained according to Fig.~\ref{Fig-Transmitter}, given by 
\begin{equation}\label{y-r}
\setlength{\abovedisplayskip}{4pt} 
\setlength{\belowdisplayskip}{4pt}
{\bf{y}} = \left( {{{\bf{F}}_N} \otimes {{\bf{I}}_M}} \right){\bf{r}} = {{\bf{H}}_{{\rm{DD}}}}{\bf{x}} + {\bf{w}},
\end{equation}
where ${{\bf{H}}_{{\rm{DD}}}} \buildrel \Delta \over = \left( {{{\bf{F}}_N} \otimes {{\bf{I}}_M}} \right){{\bf{H}}_{\rm{T}}}\left( {{\bf{F}}_N^{\rm{H}} \otimes {{\bf{I}}_M}} \right)$ represents the DD domain effective channel matrix, and ${\bf{w}} \buildrel \Delta \over = \left( {{{\bf{F}}_N} \otimes {{\bf{I}}_M}} \right){\bf{n}}$ represents the DD domain effective noise vector with the same distribution as the time domain noise.
\vspace{-0.15cm}

\subsection{Properties of the Time Domain Effective Channel Matrix}
Let us review the characteristics of the time domain effective channel matrix ${{\bf{H}}_{\rm T}}$. 
As shown in~\eqref{H-T}, 
the path delay introduces the time domain ISI.
Noticing that $0 \le l_p \le M-1$, it is clear that the interference induced by delay is restricted to adjacent time slots. 
Following the conventional OFDM setup, we partition $\bf s$ and $\bf r$ into $N$ sub-blocks, each containing $M$ samples, i.e., ${\bf{r}} = {\left[ {{\bf{r}}_0^{\rm{H}},{\bf{r}}_1^{\rm{H}}, \ldots ,{\bf{r}}_{N - 1}^{\rm{H}}} \right]^{\rm{H}}}$ and ${\bf{s}} = {\left[ {{\bf{s}}_0^{\rm{H}},{\bf{s}}_1^{\rm{H}}, \ldots ,{\bf{s}}_{N - 1}^{\rm{H}}} \right]^{\rm{H}}}$, respectively, as shown in Fig.~\ref{Fig-Time-domain-input-output}.
Thus, the $i$-th received block ${\bf r}_i$ can have interference from only the $\left(i-1\right)$-th block ${\bf s}_{i-1}$ in the time domain.
It should be noted that the first block ${\bf s}_0$ is interfered with by the last block ${\bf s}_{N-1}$ due to the appended reduced CP \cite{Viterbo-Book}.
According to this sub-block structure, ${\bf H}_{\rm T}$ can be rewritten as\par
\vspace{-0.3cm}
\begin{small}
	\begin{equation}\label{HT-Banded}
	\setlength{\abovedisplayskip}{2pt} 
	\setlength{\belowdisplayskip}{4pt}
	{{\bf{H}}_{\rm{T}}} = \left[ {\begin{array}{*{20}{c}}
		{{\bf{H}}_{\rm{T}}^{0,0}}&0&0& \cdots &0&{{\bf{H}}_{\rm{T}}^{0,1}}\\
		{{\bf{H}}_{\rm{T}}^{1,1}}&{{\bf{H}}_{\rm{T}}^{1,0}}&0& \cdots &0&0\\
		0&{{\bf{H}}_{\rm{T}}^{2,1}}&{{\bf{H}}_{\rm{T}}^{2,0}}& \cdots &0&0\\
		\vdots & \vdots & \ddots & \ddots & \vdots & \vdots \\
		0&0&0& \cdots &{{\bf{H}}_{\rm{T}}^{N - 2,0}}&0\\
		0&0&0& \cdots &{{\bf{H}}_{\rm{T}}^{N - 1,1}}&{{\bf{H}}_{\rm{T}}^{N - 1,0}}
		\end{array}} \right],
	\end{equation} 
\end{small}%
where ${{\bf{H}}_{\rm{T}}^{i,0}}$, $i=0,\ldots,N-1$, of size $M \times M$ are the diagonal blocks of ${\bf H}_{\rm T}$, given by 
\begin{equation}\label{H_T_i0}
\setlength{\abovedisplayskip}{4pt} 
{\bf{H}}_{\rm{T}}^{i,0} = \sum\limits_{p = 1}^P {{h_p}{e^{ - j2\pi \frac{{{k_p} + {\kappa _p}}}{{NM}}{l_p}}}} {\bf{\Delta }}_i^{{k_p} + {\kappa _p}}{\bf{\Pi }}_0^{{l_p}},    
\end{equation}
and ${{\bf{H}}_{\rm{T}}^{i,1}}$, $i=0,\ldots,N-1$, of size $M \times M$ are the first sub-diagonal blocks of ${\bf H}_{\rm T}$, i.e.,
\begin{equation}\label{H_T_i1}
\setlength{\abovedisplayskip}{4pt} 
\setlength{\belowdisplayskip}{4pt}
{\bf{H}}_{\rm{T}}^{i,1} = \sum\limits_{p = 1}^P {{h_p}{e^{ - j2\pi \frac{{{k_p} + {\kappa _p}}}{{NM}}{l_p}}}} {\bf{\Delta }}_i^{{k_p} + {\kappa _p}}{\bf{\Pi }}_1^{{l_p}},   
\end{equation}
representing the inter-block interference from the $\left(i-1\right)$-th transmitted block to the $i$-th received block \cite{Viterbo-Book}.
In addition, in \eqref{H_T_i0} and \eqref{H_T_i1}, ${{\bf{\Delta }}_i} = {\rm{diag}}\left\{ {{\alpha ^{iM}},{\alpha ^{iM+1}}, \ldots ,{\alpha ^{{iM}+M - 1}}} \right\}$ characterizes the Doppler effect on the $i$-th received block, while
${\bf{\Pi }}_0$ and ${\bf{\Pi }}_1$ are the \emph{forward shift} and \emph{backward shift} matrices of size $M \times M$, respectively, i.e.,\par
\vspace{-0.3cm}
\begin{small}
	\begin{align}\label{Pi-0-1}
	\setlength{\abovedisplayskip}{0pt} 
	\setlength{\belowdisplayskip}{4pt}
	{\bf{\Pi }}_0 = {\left[ {\begin{array}{*{20}{c}}
			0& \ldots &0&0\\
			1& \ldots &0&0\\
			\vdots & \ddots & \ddots & \vdots \\
			0& \ldots &1&0
			\end{array}} \right]},
	{\bf{\Pi }}_1 = {\left[ {\begin{array}{*{20}{c}}
			0& 1 &\ldots&0\\		
			\vdots & \ddots & \ddots & \vdots \\
			0& 0 &\ldots&1\\
			0& 0 &\ldots&0
			\end{array}} \right]}.
	\end{align}  
\end{small}%
Furthermore, we notice that the Doppler effect behaves like a phase term in ${{\bf{H}}_{\rm T}} $, which does not affect the channel sparsity.
The above observation suggests that the time domain effective channel matrix ${{\bf{H}}_{\rm T}} $ has a ``band-limited" structure, where each row only has limited non-zero elements{\footnote{We note that this observation holds roughly in the case of fractional delay. However, the inter-block interference may span several sub-blocks due to the insufficient delay resolution.}}.

According to the above analysis, we reformulate the block-wise input-output relation in the time domain by
\begin{equation}\label{ri}
\setlength{\abovedisplayskip}{4pt} 
\setlength{\belowdisplayskip}{4pt}
{{\bf{r}}_i} = {\bf{H}}_{\rm{T}}^{i,0}{{\bf{s}}_i} + {\bf{H}}_{\rm{T}}^{i,1}{{\bf{s}}_{{{\left( {i - 1} \right)}_N}}} + {{\bf{n}}_i},
\end{equation}
$i=0,\ldots,N-1$, where $\left(\cdot\right)_N$ denotes mod-$N$ operation. Note that both ${{\bf{r}}_i}$ and ${{{\bf{r}}_{{{\left( {i + 1} \right)}_N}}}}$ contain the information of ${\bf s}_i$ due to the path delay. Thus, it is convenient to write \par
\vspace{-0.3cm}
{\footnotesize
	\begin{align}\label{ri-ri1}
	\hspace{-1.5mm}
	\setlength{\abovedisplayskip}{4pt} 
	\setlength{\belowdisplayskip}{4pt}
	\left[\!\! {\begin{array}{*{20}{c}}
		{{{\bf{r}}_i}}\\
		{{{\bf{r}}_{{{\left( {i + 1} \right)}_N}}}}
		\end{array}} \!\!\right] &= \left[\!\! {\begin{array}{*{20}{c}}
		{{\bf{H}}_{\rm{T}}^{i,0}}\\
		{{\bf{H}}_{\rm{T}}^{{{\left( {i + 1} \right)}_N},1}}
		\end{array}} \!\!\right]{{\bf{s}}_i} + \left[\!\! {\begin{array}{*{20}{c}}
		{{\bf{H}}_{\rm{T}}^{i,1}}&{}\\
		{}&{{\bf{H}}_{\rm{T}}^{{{\left( {i + 1} \right)}_N},0}}
		\end{array}} \!\!\right]\left[\!\! {\begin{array}{*{20}{c}}
		{{{\bf{s}}_{ {\left(i-1\right)}_N }}}\\
		{{{\bf{s}}_{{{\left( {i + 1} \right)}_N}}}}
		\end{array}} \!\!\right] \notag\\
	&\quad+ \left[\!\! {\begin{array}{*{20}{c}}
		{{{\bf{n}}_i}}\\
		{{{\bf{n}}_{{{\left( {i + 1} \right)}_N}}}}
		\end{array}} \!\!\right],
	\end{align}	
}\vspace{-0.1cm}%
which can be further written as
\begin{equation}\label{tilde-r}
\setlength{\abovedisplayskip}{4pt} 
\setlength{\belowdisplayskip}{4pt}
{{\bf{\tilde r}}_i} = {\bf{H}}_{\rm{A}}^i{{\bf{s}}_i} + {\bf{H}}_{\rm{B}}^i{{\bf{\tilde s}}_i} + {{\bf{\tilde n}}_i}.
\end{equation}
In \eqref{tilde-r}, ${{\bf{\tilde r}}_i}\!\! \in\!\!{\mathbb C}^{2M \times 1}$ is the observation at the receiver side corresponding to ${\bf{s}}_i$; ${{\bf{H}}_{\rm{A}}^i} \in {\mathbb C}^{2M \times M}$ is the effective observation matrix, characterizing the interference pattern related to ${\bf s}_i$; ${{\bf{H}}_{\rm{B}}^i} \in {\mathbb C}^{2M \times 2M}$ is the effective interference matrix, characterizing the additional interference from other transmitted sub-blocks; ${{\bf{\tilde s}}_i}\in {\mathbb C}^{2M \times 1}$ is the interfering signal vector; and ${{\bf{\tilde n}}_i} \in {\mathbb C}^{2M \times 1} $ is the considered noise vector.

The above discussion naturally motivates us to design a time domain estimator/detector that exploits the effective channel matrix structure, and this is presented in the following section.

\section{Cross-Domain Iterative Detection for OTFS Modulation via Delay-Doppler Decoupling}
\begin{figure}[htp]
	\centering
	\includegraphics[width=3.5in,height=0.93in]{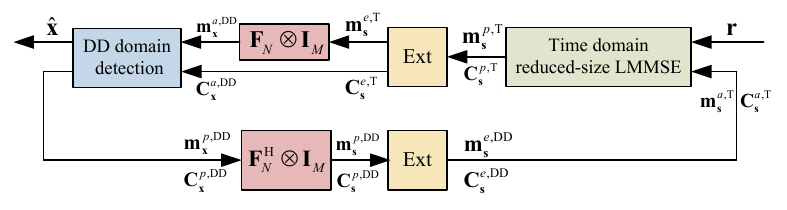}
	\vspace{-0.6cm}
	\caption{The block diagram of the considered cross-domain iterative receiver with the time domain reduced-size LMMSE.}\label{Fig-Receiver}
	\vspace{-0.5cm}
\end{figure}
Based on \eqref{tilde-r}, we propose to apply a reduced-size LMMSE estimator for eliminating the delay interference, which has a much lower complexity compared to the full-size LMMSE estimator adopted in~\cite{S.Li-2022TWC-Crossdomain}.
However, such estimator cannot fully minimize the effect of Doppler. In~\cite{S.Li-2022TWC-Crossdomain}, the authors have shown that the cross-domain iteration can effectively detect the OTFS signal by iteratively exchanging the extrinsic information between the time domain and the DD domain due to the fact that the time and Doppler are a pair of Fourier dual. Following this idea, we adopt the cross-domain iteration to minimize the Doppler interference as will be discussed later.

\subsection{Reduced-Size LMMSE Estimator in the Time Domain}
Based on \eqref{tilde-r}, the block-wise LMMSE estimation matrix ${\bf W}_{\rm{MMSE}}^i$ for the $i$-th transmitted block ${\bf s}_i$ can be obtained as
\begin{align}\label{W-i}
\setlength{\abovedisplayskip}{4pt} 
\setlength{\belowdisplayskip}{4pt}
{\bf{W}}_{{\rm{MMSE}}}^i &= {\bf{C}}_{{{\bf{s}}_i}}^{a,{\rm{T}}}{\left( {{\bf{H}}_{\rm{A}}^i} \right)^{\rm{H}}}\left( {{\bf{H}}_{\rm{A}}^i{\bf{C}}_{{{\bf{s}}_i}}^{a,{\rm{T}}}{{\left( {{\bf{H}}_{\rm{A}}^i} \right)}^{\rm{H}}}} \right. \notag\\
&\;\quad{\left. { + {\bf{H}}_{\rm{B}}^i{\bf{C}}_{{{{\bf{\tilde s}}}_i}}^{a,{\rm{T}}}{{\left( {{\bf{H}}_{\rm{B}}^i} \right)}^{\rm{H}}} + {N_0}{{\bf{I}}_{2M}}} \right)^{ - 1}},
\end{align}
where ${\bf{C}}_{{{\bf{s}}_i}}^{a,{\rm{T}}}$ and ${\bf{C}}_{{{\bf{\tilde s}}_i}}^{a,{\rm{T}}}$ are the \emph{a priori} covariance matrices of ${{\bf{s}}_i}$ and ${{\bf{\tilde s}}_i}$ and initialized as ${\bf I}_{M}$ and ${\bf I}_{2M}$ for the first iteration, respectively. Furthermore, the \emph{a posteriori} estimation output ${\bf{m}}_{{{\bf{s}}_i}}^{p,{\rm{T}}}$ of ${\bf s}_{i}$ is given by
\begin{equation}\label{M-pt-si}
\hspace{-1.5mm}
\setlength{\abovedisplayskip}{4pt} 
\setlength{\belowdisplayskip}{4pt}
{\bf{m}}_{{{\bf{s}}_i}}^{p,{\rm{T}}} = {\bf{m}}_{{{\bf{s}}_i}}^{a,{\rm{T}}} + {\bf{W}}_{{\rm{MMSE}}}^i\left( {{{{\bf{\tilde r}}}_i} - {\bf{H}}_{\rm{B}}^i{\bf{m}}_{{{{\bf{\tilde s}}}_i}}^{a,{\rm{T}}} - {\bf{H}}_{\rm{A}}^i{\bf{m}}_{{{\bf{s}}_i}}^{a,{\rm{T}}}} \right),
\end{equation}
where ${\bf{m}}_{{{\bf{s}}_i}}^{a,{\rm{T}}}\!$ and ${\bf{m}}_{{{\bf{\tilde s}}_i}}^{a,{\rm{T}}}\!$ are the \emph{a priori} mean vectors of ${{\bf{s}}_i}\!$ and ${{\bf{\tilde s}}_i}$ with sizes $M \times 1$ and $2M \times 1$, respectively. Note that the above LMMSE estimator applies the successive interference cancellation (SIC) to eliminate the interference from~${{\bf{\tilde s}}_i}$ according to the \emph{a priori} information from the previous iteration.

The \emph{a posteriori} covariance matrix 
${\bf{C}}_{{{\bf{s}}_i}}^{p,{\rm{T}}}$ of ${{\bf{s}}_i}$ is given by
\begin{equation}\label{C-pt-si}
\setlength{\abovedisplayskip}{4pt} 
\setlength{\belowdisplayskip}{4pt}
{\bf{C}}_{{{\bf{s}}_i}}^{p,{\rm{T}}} = {\bf{C}}_{{{\bf{s}}_i}}^{a,{\rm{T}}} - {\bf{W}}_{{\rm{MMSE}}}^i{\bf{H}}_{\rm{A}}^i{\bf{C}}_{{{\bf{s}}_i}}^{a,{\rm{T}}}.
\end{equation}
Note that ${\bf{C}}_{{{\bf{s}}_i}}^{a,{\rm{T}}}\!\!$ should be a diagonal matrix due to the independent and identically distributed (i.i.d.) assumption of the transmitted symbols. As such, we discard the non-diagonal entries of ${\bf{C}}_{{{\bf{s}}_i}}^{p,{\rm{T}}}$ (treated as zeros) for further processing~\cite{S.Li-2022TWC-Crossdomain}. The extrinsic covariance matrix ${\bf{C}}_{{{\bf{s}}_i}}^{e,{\rm{T}}}$ and mean ${\bf{m}}_{{{\bf{s}}_i}}^{e,{\rm{T}}}$ associated with ${\bf s}_i$ from the LMMSE estimation can be written as
\begin{equation}\label{C-et-si}
\setlength{\abovedisplayskip}{4pt} 
\setlength{\belowdisplayskip}{4pt}
{\bf{C}}_{{{\bf{s}}_i}}^{e,{\rm{T}}} = {\left( {{{\left( {{\bf{C}}_{{{\bf{s}}_i}}^{p,{\rm{T}}}} \right)}^{ - 1}} - {{\left( {{\bf{C}}_{{{\bf{s}}_i}}^{a,{\rm{T}}}} \right)}^{{\rm{ - 1}}}}} \right)^{ - 1}}
\end{equation}
and
\begin{equation}\label{M-et-si}
\setlength{\abovedisplayskip}{4pt} 
\setlength{\belowdisplayskip}{4pt}
{\bf{m}}_{{{\bf{s}}_i}}^{e,{\rm{T}}} = {\bf{C}}_{{{\bf{s}}_i}}^{e,{\rm{T}}}\left( {{{\left( {{\bf{C}}_{{{\bf{s}}_i}}^{p,{\rm{T}}}} \right)}^{ - 1}}{\bf{m}}_{{{\bf{s}}_i}}^{p,{\rm{T}}} - {{\left( {{\bf{C}}_{{{\bf{s}}_i}}^{a,{\rm{T}}}} \right)}^{ - 1}}{\bf{m}}_{{{\bf{s}}_i}}^{a,{\rm{T}}}} \right),
\end{equation}
respectively.

\subsection{Cross-Domain Iterative Detection for OTFS}
The extrinsic information ${\bf{C}}_{{{\bf{s}}_i}}^{e,{\rm{T}}}$ and  ${\bf{m}}_{{{\bf{s}}_i}}^{e,{\rm{T}}}$ obtained from the LMMSE estimation is passed to the DD domain as shown in Fig.~\ref{Fig-Receiver}. 
According to the relationship between the DD domain and time domain, the DD domain \emph{a priori} information, ${\bf{C}}_{\bf{x}}^{a,{\rm{DD}}}$ and ${\bf{m}}_{\bf{x}}^{a,{\rm{DD}}}$, are given by \cite{S.Li-2022TWC-Crossdomain}
\begin{equation}\label{C-ad-s}
\setlength{\abovedisplayskip}{4pt} 
\setlength{\belowdisplayskip}{4pt}
{\bf{C}}_{\bf{x}}^{a,{\rm{DD}}} = {\bf{C}}_{\bf{s}}^{e,{\rm{T}}},
\end{equation}
and
\begin{equation}\label{M-ad-x}
\setlength{\abovedisplayskip}{4pt} 
\setlength{\belowdisplayskip}{4pt}
{\bf{m}}_{\bf{x}}^{a,{\rm{DD}}}= \left( {{{\bf{F}}_N} \otimes {{\bf{I}}_M}} \right){\bf{m}}_{\bf{s}}^{e,{\rm{T}}},
\end{equation}
where ${\bf{C}}_{\bf{s}}^{e,{\rm{T}}} = {\rm{diag}}\left\{ {{\rm{diag}}\left\{ {{\bf{C}}_{{{\bf{s}}_0}}^{e,{\rm{T}}}} \right\}, \ldots ,{\rm{diag}}\left\{ {{\bf{C}}_{{{\bf{s}}_{N - 1}}}^{e,{\rm{T}}}} \right\}} \right\}$ and ${\bf{m}}_{\bf{s}}^{e,{\rm{T}}} \!\!=\! {\left[ {{{\left( {{\bf{m}}_{{{\bf{s}}_0}}^{e,{\rm{T}}}} \right)}^{\rm{T}}}, \ldots ,{{\left( {{\bf{m}}_{{{\bf{s}}_{N - 1}}}^{e,{\rm{T}}}} \right)}^{\rm{T}}}} \right]^{\rm{T}}}\!\!\!$ are the extrinsic covariance matrix and mean of $\bf s$, respectively.

In the DD domain, the detection can be conducted in a simple symbol-by-symbol manner, e.g., Algorithm 2 in~\cite{S.Li-2022TWC-Crossdomain}, where the corresponding \emph{a posteriori} mean ${\bf{m}}_{\bf{x}}^{p,{\rm{DD}}}\!$ and covariance matrix ${\bf{C}}_{\bf{x}}^{p,{\rm{DD}}}\!$ of the DD domain OTFS symbol $\bf x$ are then passed back to the time domain for calculating the extrinsic information{\footnote{As discussed in~\cite{S.Li-2022TWC-Crossdomain}, the extrinsic information cannot be directly calculated in the DD domain because of the symbol-by-symbol detection.}}. Based on \eqref{s=x}, the corresponding \emph{a posteriori} mean ${\bf{m}}_{\bf{s}}^{p,{\rm{DD}}}\!$ and covariance matrix ${\bf{C}}_{\bf{s}}^{p,{\rm{DD}}}\!$ of $\bf s$ are given by
\begin{equation}\label{M-pd-s}
\setlength{\abovedisplayskip}{4pt} 
\setlength{\belowdisplayskip}{4pt}
{\bf{m}}_{\bf{s}}^{p,{\rm{DD}}} = \left( {{\bf{F}}_N^{\rm{H}} \otimes {{\bf{I}}_M}} \right){\bf{m}}_{\bf{x}}^{p,{\rm{DD}}},
\end{equation}
and
\begin{equation}\label{C-pd-s}
\setlength{\abovedisplayskip}{4pt} 
\setlength{\belowdisplayskip}{4pt}
{\bf{C}}_{\bf{s}}^{p,{\rm{DD}}} = \left( {{\bf{F}}_N^{\rm{H}} \otimes {{\bf{I}}_M}} \right){\bf{C}}_{\bf{x}}^{p,{\rm{DD}}}\left( {{{\bf{F}}_N} \otimes {{\bf{I}}_M}} \right),
\end{equation}
respectively.
Then, the extrinsic information of $\bf s$ in terms of the covariance matrix and mean can be obtained as~\cite{S.Li-2022TWC-Crossdomain}
\begin{equation}\label{C-ed-s}
\setlength{\abovedisplayskip}{4pt} 
\setlength{\belowdisplayskip}{4pt}
{\bf{C}}_{\bf{s}}^{e,{\rm{DD}}} = {\left( {{{\left( {{\bf{C}}_{\bf{s}}^{p,{\rm{DD}}}} \right)}^{ - 1}} - {{\left( {{\bf{C}}_{\bf{s}}^{a,{\rm{DD}}}} \right)}^{{\rm{ - 1}}}}} \right)^{ - 1}},
\end{equation}
and
\begin{equation}\label{M-ed-s}
\hspace{-2mm}
\setlength{\abovedisplayskip}{4pt} 
\setlength{\belowdisplayskip}{4pt}
{\bf{m}}_{\bf{s}}^{e,{\rm{DD}}} \!=\! {\bf{C}}_{\bf{s}}^{e,{\rm{DD}}}\!\left(\! {{{\left( {{\bf{C}}_{\bf{s}}^{p,{\rm{DD}}}} \right)}^{ \!- 1}}{\bf{m}}_{\bf{s}}^{p,{\rm{DD}}} \!-\! {{\left( {{\bf{C}}_{\bf{s}}^{a,{\rm{DD}}}} \right)}^{\! - 1}}{\bf{m}}_{\bf{s}}^{e,{\rm{T}}}} \! \right),
\end{equation}
respectively.
Next, the extrinsic information is fed back to the time domain LMMSE estimator for the coming iteration. Specifically, the \emph{a priori} covariance matrix and mean of $\bf s$ are updated to ${\bf{C}}_{\bf{s}}^{a,{\rm{T}}}={\bf{C}}_{\bf{s}}^{e,{\rm{DD}}}$ and ${\bf{m}}_{\bf{s}}^{a,{\rm{T}}}={\bf{m}}_{\bf{s}}^{e,{\rm{DD}}}$.

\section{Performance Analysis}
\subsection{Performance Analysis by State Evolution}
In this subsection, we will characterize the asymptotic MSE performance of the proposed detection scheme by state evolution.
Without loss of generality, we define the average \emph{a priori} variance of the inputs to the time domain estimator and the DD domain estimator in the $l$-th iteration as
\begin{equation}
\hspace{-1.5mm}
\setlength{\abovedisplayskip}{4pt} 
\setlength{\belowdisplayskip}{4pt}
v_s^{a,{\rm{T}}}\left( l \right)\mathop  = \limits^\Delta  {\mathbb{E}}\left[ {C_{\bf{s}}^{e,{\rm{DD}}}\left[ {k,k} \right]} \right] = \mathop {\lim }\limits_{MN \to \infty } \frac{1}{{MN}}{\rm{Tr}}\left( {{\bf{C}}_{\bf{s}}^{e,{\rm{DD}}}} \right),
\end{equation}
\begin{equation}
\setlength{\abovedisplayskip}{4pt} 
\setlength{\belowdisplayskip}{4pt}
v_s^{a,{\rm{DD}}}\left( l \right)\mathop  = \limits^\Delta  {\mathbb{E}}\left[ {C_{\bf{s}}^{e,{\rm{T}}}\left[ {k,k} \right]} \right] = \mathop {\lim }\limits_{MN \to \infty } \frac{1}{{MN}}{\rm{Tr}}\left( {{\bf{C}}_{\bf{s}}^{e,{\rm{T}}}} \right),
\end{equation}
where the expectation is with respect to the symbol index $k$.
The two states can also be viewed as the asymptotically average MSEs of inputs in the $l$-th iteration. 
Assume that the main diagonal entries of  ${{\bf{C}}_{\bf{s}}^{a,{\rm{T}}}}$ and ${{\bf{C}}_{\bf{x}}^{a,{\rm{DD}}}}$ are of the same value as $v_s^{a,{\rm{T}}}\left( l \right)$ and $v_s^{a,{\rm{DD}}}\left( l \right)$, respectively, for the $l$-th iteration, $v_s^{a,{\rm{DD}}}\left( l \right)$ can be represented as~\cite{S.Li-2022TWC-Crossdomain}
\begin{equation}\label{v-s-a-DD-l}
\setlength{\abovedisplayskip}{4pt} 
\setlength{\belowdisplayskip}{4pt}
v_s^{a,{\rm{DD}}}\left( l \right) = \frac{1}{{\frac{1}{{v_s^{p,{\rm{T}}}\left( l \right)}} - \frac{1}{{v_s^{a,{\rm{T}}}\left( l \right)}}}},
\end{equation}
according to \eqref{C-et-si}. Based on~\eqref{C-pt-si}, the average \emph{a posteriori} variance of ${\bf{C}}_{{{\bf{s}} }}^{p,{\rm{T}}}$ can be obtained as \par
\begin{small}
	\begin{align}\label{v-s-pT}
	\hspace{-2.5mm}
	v_s^{p,{\rm{T}}}\!\left( l \right) &=\! v_s^{a,{\rm{T}}}\!\left( l \right) \!-\! \frac{{{{\left( {v_s^{a,{\rm{T}}}\left( l \right)} \right)}^2}}}{{MN}} \! \times \! \sum\limits_{i = 0}^{N - 1} {{\rm{Tr}}} \left\{\! {{{\left( {{\bf{H}}_{\rm{A}}^i} \right)}^{\!\rm{H}}}\!\left( {v_s^{a,{\rm{T}}}\!\left( l \right)} \right.}  {\bf{H}}_{\rm{A}}^i\right. \notag \\
	&\quad\left. {{{\left. {{{\left( {{\bf{H}}_{\rm{A}}^i} \right)}^{\!\rm{H}}} \!+\! v_s^{a,{\rm{T}}}\!\left( l \right){\bf{H}}_{\rm{B}}^i{{\left( {{\bf{H}}_{\rm{B}}^i} \right)}^{\!\rm{H}}} \!+\! {N_0}{{\bf{I}}_{2M}}} \right)}^{ - 1}}{\bf{H}}_{\rm{A}}^i} \!\right\}.
	\end{align}
\end{small}%
According to \cite{S.Li-2022TWC-Crossdomain}, the update of the state $v_s^{a,{\rm{T}}}\left( l+1 \right)$ can be given by
\begin{equation}\label{v-s-aT-l+1}
\setlength{\abovedisplayskip}{4pt} 
\setlength{\belowdisplayskip}{4pt}
v_s^{a,{\rm{T}}}\left( {l + 1} \right) = \frac{1}{{\frac{1}{{v_s^{p,{\rm{DD}}}\left( l \right)}} - \frac{1}{{v_s^{a,{\rm{DD}}}\left( l \right)}}}},
\end{equation}
where $v_s^{p,{\rm{DD}}}\left( l \right)\! \buildrel \Delta \over =\! \!\!\mathop {\lim }\limits_{MN \to \infty } \frac{1}{{MN}}{\rm{Tr}}\left( {{\bf{C}}_{\bf{x}}^{p,{\rm{DD}}}} \right)$. By iteratively updating the MSE state according to~\eqref{v-s-a-DD-l} and~\eqref{v-s-aT-l+1}, the state evolution can then be derived.
Note that the above derivation is heuristic due to the utilization of the Fourier transform. Rigorous analysis of error performance may be discussed in future work.

In addition to the derived state evolution, we further apply bounding techniques to discuss the insights of the proposed scheme. Note that the proposed scheme adopts the SIC for time domain estimation. Therefore, depending on whether SIC can fully eliminate the interference, the MSE performance can be bounded by applying the treating interference as noise (TIN) strategy (corresponding to the worst-case scenario) and the genie-aided strategy (corresponding to the best-case scenario). Due to the space limitation, we could not provide the full details of these two bounds. However, we point out that these two bounds can be derived by modifying the corresponding covariance matrix of the interference terms $v_s^{a,{\rm{T}}}\left( l \right){\bf{H}}_{\rm{B}}^i{{\left( {{\bf{H}}_{\rm{B}}^i} \right)}^{\rm{H}}}$ in~\eqref{v-s-pT}, and further details of the bounds will be presented in our journal paper.

It should be noted that both TIN and genie-aided bounds are of theoretical significance, as they together indicate whether the residual interference in the time domain can be minimized by the cross-domain iteration. As we will demonstrate in the numerical results part, the TIN bound and genie-aided bound will converge to each other after a sufficient number of cross-domain iterations. This suggests that the adopted reduced-size LMMSE estimator aligns well with the mechanism of the cross-domain iterative detection, where only the interference caused by the delay needs to be considered for the time domain estimation, while the interference caused by Doppler can be resolved naturally by the cross-domain iteration.

\subsection{Complexity Analysis}
The complexity of the cross-domain iterative detection has been reported in~\cite{S.Li-2022TWC-Crossdomain}. In comparison, the proposed scheme reduces the complexity by using a reduced-size LMMSE estimator, whose complexity is in the order of ${\cal O} \left( \left(2M\right)^3N \right)$. Furthermore, the complexity of the cross-domain iteration and the DD domain symbol-by-symbol detection is ${\cal O} \left( MN{\rm{log}}N \right)$ and ${\cal O} \left( MN \right)$, respectively. 
Therefore, the proposed scheme in total requires ${\cal O} \left(8M^3N + MN{\rm{log}}N + MN \right)$ complexity per iteration, which is much less compared to ${\cal O} \left(M^3N^3 + MN{\rm{log}}N + MN \right)$ required in~\cite{S.Li-2022TWC-Crossdomain}.


\begin{figure}[htp]
	\centering
	\vspace{-0.25cm}
	\includegraphics[width=3in,height=2.3in]{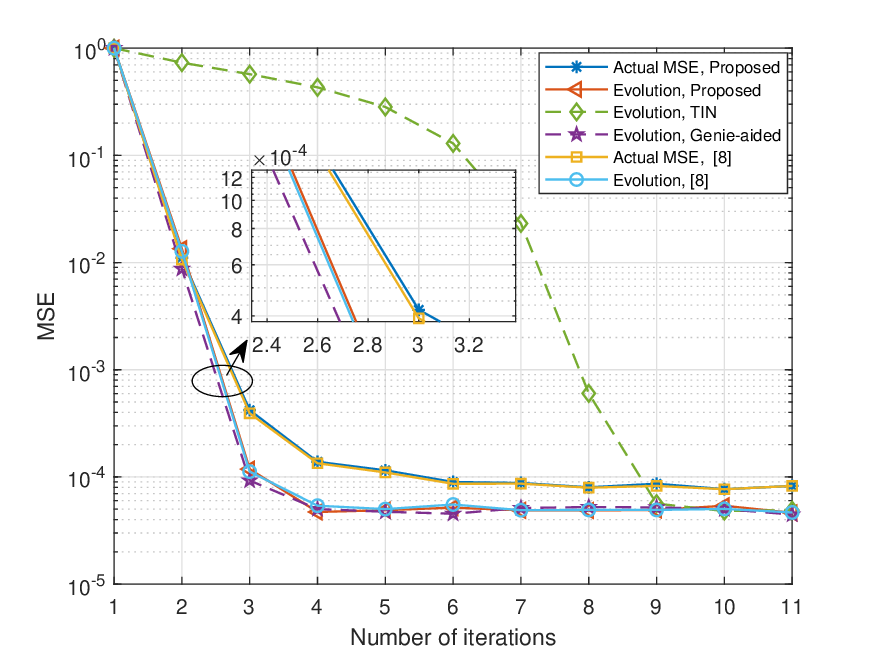}
	\vspace{-0.3cm}
	\caption{Comparison of time domain state (MSE) evolution performance of the proposed scheme and the scheme in \cite{S.Li-2022TWC-Crossdomain}, as well as the derived bounds.}\label{Figure-Compare-MSE-P4-Fractional}
\end{figure}
\begin{figure}[htp]
	\centering
	\vspace{-0.5cm}
	\includegraphics[width=3in,height=2.3in]{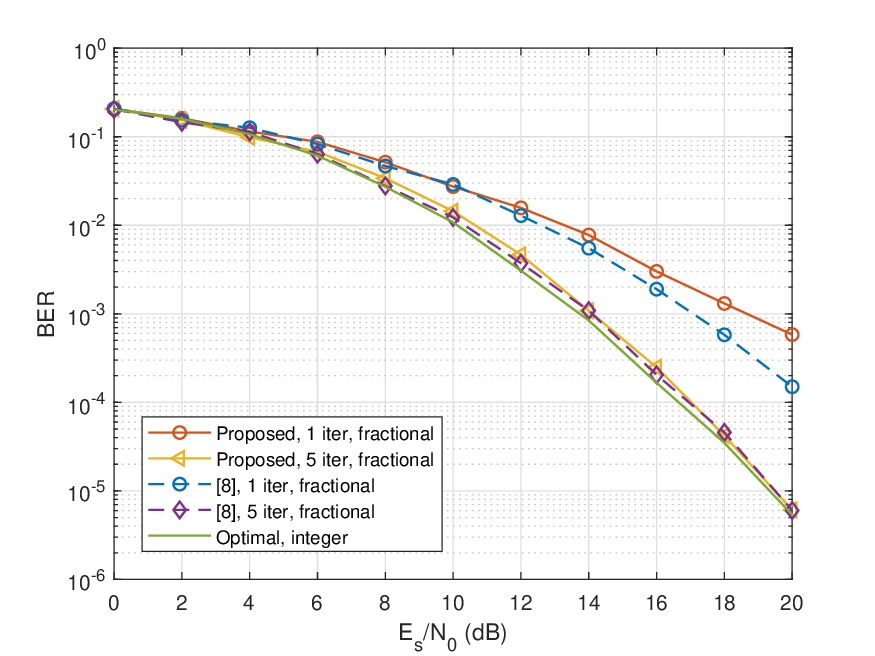}
	\vspace{-0.3cm}
	\caption{Comparison of BER performance of the proposed scheme, the scheme in \cite{S.Li-2022TWC-Crossdomain}, and the optimal detection in \cite{S.Li-2021TVT-MAP-PIC}.}\label{Figure-Compare-Error-P4-Fractional}
	\vspace{-0.4cm}
\end{figure}

\section{Numerical Results}
\vspace{-0.1cm}
We present numerical results of the proposed scheme in this section. {As an example, consider} $P=4$, $M=64$, $N=32$, QPSK. The maximum delay and Doppler index are $10$ and $5$.

The state (MSE) evolution performance of the proposed scheme at a signal-to-noise ratio (SNR) ${E_s}/{N_0}=12$~dB, is given in Fig.~\ref{Figure-Compare-MSE-P4-Fractional}, where both the actual MSE, the MSE evolution results, and the derived TIN and genie-aided bounds are presented in comparison to the results obtained from~\cite{S.Li-2022TWC-Crossdomain}\setlength{\skip\footins}{0.2cm}{\footnote{In Fig.~\ref{Figure-Compare-MSE-P4-Fractional}, the delay indices are $\left[0,8,4,6\right]$, the Doppler indices are $\left[4.82,-3.23,1.38,-2.47\right]$, and the channel coefficients are $\left[-0.02-0.09i,   0.40+0.73i,0.03+0.45i,0.15-0.43i\right]$, respectively.}}.
As shown in Fig.~\ref{Figure-Compare-MSE-P4-Fractional}, our derived state evolution provides a good prediction of the actual MSE performance, where both the actual MSE and the MSE derived from the state evolution decrease first and then saturate at MSEs around $9 \times {10^{-5}}$, after sufficient iterations.
Furthermore, we can observe that the derived state evolution matches perfectly with both the TIN and genie-aided bounds, which verifies the correctness of our derivation.
Finally, we notice that the proposed scheme only exhibits marginal MSE loss compared to the scheme in~\cite{S.Li-2022TWC-Crossdomain} in early iterations both numerically and theoretically, and this loss becomes negligible with an increased number of iterations. 

In Fig. \ref{Figure-Compare-Error-P4-Fractional}, we compare the bit error rate (BER) performance of the proposed scheme and the scheme in \cite{S.Li-2022TWC-Crossdomain}.
As observed from the figure, the proposed scheme suffers from a noticeable performance degradation with one iteration compared to the scheme in~\cite{S.Li-2022TWC-Crossdomain} at high SNRs, due to the imperfect SIC adopted in the scheme in early iterations. However, we notice that the proposed scheme shows roughly the same performance as the scheme in~\cite{S.Li-2022TWC-Crossdomain} with $5$ iterations, and both their results converge to the optimal performance obtained by using the MP algorithm~\cite{S.Li-2021TVT-MAP-PIC} with only integer delay and Doppler indices. Therefore, we observe that the proposed scheme enjoys a near-optimal performance with reduced complexity.

\section{Conclusion}
\vspace{-0.10cm}
In this paper, we proposed a novel reduced-complexity cross-domain iterative detection for OTFS transmissions. By utilizing the ``bandlimitedness'' of the time domain effective channel matrix, the reduced-size time domain LMMSE is performed in a block-wise manner, where we showed that the potential performance degradation due to this reduced-size estimation can be well compensated by using the cross-domain framework. The state evolution and theoretical bounds were also provided.
Finally, numerical results showed that the proposed scheme achieves a near-optimal performance.

\vspace{-0.1cm}
\section*{Acknowledgment}
\vspace{-0.1cm}
This work was supported in part by the National Key R\&D Program of China under Grant 2021YFA1000500, and in part by the Federal Ministry of Education and Research of Germany in the program of ``Souver{\"a}n. Digital. Vernetzt.'' Joint project 6G-RIC, project identification number: 16KISK030.
%


\end{document}